\documentclass{elsart}
\usepackage{epsfig}
\usepackage{amssymb}
\begin{document}
\begin{frontmatter}

%\title{Solar proton detection over 40 GeV in a Class M solar flare}
\title{Did the 2000 November 8 solar flare accelerate protons to ¡Ý40 GeV?}
\author{R. G. Wang, L. K. Ding, Y. Q. Ma, X. H. Ma, Q. Q. Zhu,}
\author{C. G. Yang, H. H. Kuang, Z. Q. Yu, Z. G. Yao, Y. P. Xu}
\address{Institute of High Energy Physics, Chinese Academy
of Sciences, Beijing 100049, China}

\begin{abstract}
It has been reported that a 5.7$\sigma$ directional muon excess
coincident with the 2000 July 14 solar flare was registered by the
L3 precision muon spectrometer [Ruiguang Wang, Astroparticle Phys., 31(2009) 149].
Using a same analysis method and similar criteria of event selection, we have analyzed the L3 precision muon spectrometer data during November 2000. The result shows that a 4.7$\sigma$ muon excess appeared at a time coincident with the solar flare of 8 of November 2000. This muon excess corresponds to above 40 GeV primary protons which came from a sky cell of solid angle 0.048 sr. The probability of being a background fluctuation is estimated to be about 0.1\%. It has been convinced that solar protons could be accelerated to tens of GeV in those Class X solar flares which usually arose solar cosmic ray ground level enhancement (GLE) events. However, whether a Class M solar flare like the non-GLE event of 8 November 2000 may also accelerate solar protons to such high energies? It is interesting and noteworthy.

\end{abstract}
\end{frontmatter}

\section{Introduction}
It is known that solar energetic particles are accelerated during intense solar high energy process
which usually accompanied by solar flares (SFs) and/or coronal mass ejections (CMEs). When the
fluxe of solar protons with energies above 10 MeV exceeds 10 pfu (1 pfu = 1 proton $(cm^{2}\cdot sec\cdot ster)^{-1}$ ),
the event is referred to as a solar proton event (SPE). From 1976 to now, more than 265 SPEs
have been observed by spacecraft-based detectors\cite{web1}.
Those SPEs, in which if solar energetic
particles with energies above several hundred MeV produce particles cascade at the top of the
atmosphere and give rise to a available flux increase of cosmic rays on the surface of the Earth,
are also called cosmic ray ground level enhancements (GLEs).
Up to now, 71 GLEs have been recorded mainly by the worldwide network of neutron monitors
(NMs)\cite{weboulu} since the first GLE observation
in 1942\cite{Forbush46}.

The worldwide NMs detect the fluxes of secondary neutrons produced by incident protons
coming from all acceptance at ground level and show the energy threshold of incident protons with
the local geomagnetic rigidities. Compared to NMs, some directional detectors (such as muon
telescopes) operating at higher energies should be better suited for detecting higher energy solar
proton beams in big solar flares. Since energy thresholds are typically fixed by the instrument
design and their atmospheric or underground depth, these instruments can register events, in
principle, with energies up to or beyond 100 GeV \cite{Chilingarian}.
Unfortunately, few such muon detectors is
in operation or running for a long period in the world.

In recent years, utilizing the technique of particle trajectory tracing \cite{Cramp1997} a GLE can be
modelled from different NMs' data with an advanced model
of the magnetospheric magnetic field \cite{Tsyganenko1989,Tsyganenko2002}. The solar proton beam approaching the Earth can be described in simulations \cite{Duldig2003,Bieber2002,Bombardieri2006}.
Many such studies showed that the arrival direction of relative solar protons in big flares is often
anisotropic, sometimes quite anisotropic, and these protons often follow a steep spectrum.

Lots of investigations of SPEs and GLEs have been reported in many typical papers, such as
these works \cite{Cliver1982,Cliver2006,wangapp2,wangasr1,wangasr2,cane10,Firoz11a,Firoz11b,Firoz12,Aschwanden}.
Measurements of the GLEs have indicated that the Sun could accelerate protons up to tens of GeV in energy \cite{Parker1957,navia05,wangapp1,l3sf06,wang09}.
In contrast, the information on solar relativistic protons produced in those SPEs of non-GLEs is still scarce.
It is very interest to know whether there are still high energy solar proton beams and how high energies the solar protons can be accelerated in such big SPEs of non-GLE.
In 1971 a positive correlation was obtained between a significant muon intensity increase and a specific solar flare by a narrow-angle
telescope located at a underground mine in Colorado \cite{nat71},
although no response of NMs to this
event. It is the first experiment evidence for solar particle production in above $\sim$75 GeV energy
region in a solar flare of class M8. After that we have not seen any significant report on this subject.

The L3+C experiment could measure the momentum and direction of cosmic ray muons \cite{Adriani2002}
with the precision muon chamber
of the L3 spectrometer \cite{Adeva1990}.
Its typical superiorities should be high directional resolution, high momentum
resolution, low momentum threshold and a large sensitive volume. Its running periods
(1999 - 2000) were just covering the peak years of the solar cycle 23. These factors offer a great opportunity
for us to search for high energy solar protons in SPEs and to try to answer the questions
discussed above.

Using L3 precision muon spectrometry, a muon excess correlated with the GLE of 14 July 2000
was obtained from a small sky cell of a solid angle of 0.046 sr \cite{l3sf06,wang09},
corresponding to primary
solar protons from 40 GeV to 100 GeV. Exception the GLE of July 14, a biggest SPE happened
on November 8 of this year. Adopting the same analysis technique for the GLE of 14 July 2000,
we also analyzed the data of 8-9 November 2000 and found a new muon excess. In this paper we will present the
data analysis and the results for this event. After a brief introduction of
this event and our experiment in the next two sections, we will mainly explain the data analysis technique
and the results in Section 4, following some discussions and a conclusion in the last two sections.

\section{The solar proton event of 8 November 2000}
The SPE of 8 November 2000 was associated with an M7.4/3F class parent solar flare produced in
the optical coordinates N10W77. Several small flares in the regions NOAA9212, NOAA9213 and
NOAA9218, gave rise to this strong M7.4 flare at 23:38 UT and triggered a strong solar storm of
high-energy particles \cite{web3}.
The X-ray flare, lasting from 22:42 UT to 00:05 UT of November 9
with a peak at 23:28 UT, was accompanied by a fast partial halo CME at 23:06 UT \cite{web5}.
The CME
speed was about 1738 km/s. It was the second largest SPE ever occurred in 2000s. A
type IV radio burst happened at 23:45 UT \cite{web6},
designating
the starting of high energy phenomena in the flare and being thought to be close to the time of
relativistic proton acceleration \cite{Cliver1982}.
Soon after that, a strong solar storm of high-energy particles
was triggered. The satellite-borne detectors, both GOES-8 and GOES-10, observed a rapid increase
of proton fluxes with energies larger than 10 MeV, 50 MeV and 100 MeV, respectively. At about
00:10 UT of 9th the flux of protons with energy up to 500 MeV peaked in about 3 orders high.
The Solar-Terrestrial activity chart \cite{web7}
shows solar wind exceeded about 900 km/s
about two days later. It is interesting to note that no NMs have significant response to this event.

\section{The L3+C Experiment}
\begin{figure}
\centering
\includegraphics[angle=0,scale=.50]{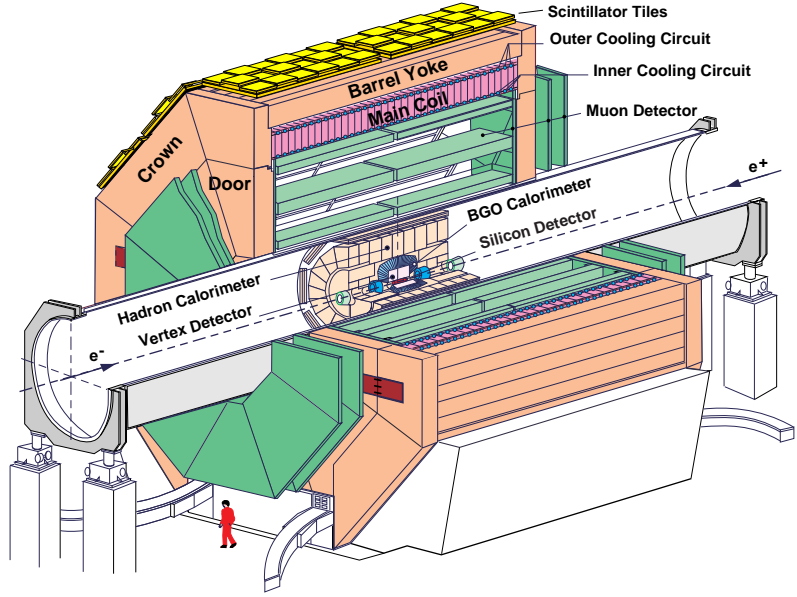}
\caption{The L3 spectrometer. Only the muon
detectors, the magnet and the scintillator tiles were used in this
experiment.}\label{l3c}
\end{figure}

The L3+Cosmics (L3+C) detector \cite{Adriani2002}
combines the high precision muon drift chambers
of the L3 spectrometer with an air shower array on the surface.
As shown in Figure~\ref{l3c}, only the muon detectors, the magnet and the scintillator tiles of the L3 spectrometer were used to measure cosmic rays. The muon drift chamber, installed in a 1000 $m^{3}$ magnetic field of 0.5 T, shows an octant shape in the plane perpendicular to the beam
(11 m in width and 11 m in height) and a square shape in the plane along the beam (11 m in
length). The maximum geometrical acceptance is $\sim200 m^2 sr$, covering a zenith
angle range from 0$^\circ$ to $\sim
60^\circ$.

The detector is located shallow underground near Geneva (6.02$^\circ$E, 46.25$^\circ$N) at an altitude of 450 m above
sea level, where the vertical geomagnetic rigidity cutoff is $\sim$5 GV. The approximately 30 m molasse above the detector provides
a 15 GeV cutoff for the muon energy, corresponding to primary proton energies above 40 GeV.
In order to independently observe cosmic ray events a timing detector, composed of 202 $m^{2}$ of
plastic scintillators, was installed on top of the magnet, and a separate trigger and DAQ system
were used for the data taking of cosmic ray events.

Although a independent data taking system was set up for cosmic ray event register, much
high background still existed when LEP positron-electron collisor was operating. Fortunately, the
collision experiment has been stopped after the end of October 2000. So the subsequent muon
data taken by muon drift chambers had very low background, which is very crucial for analysis of
the SPE of 8 November 2000.

\section{Data analysis and results}
This analysis is to search for possible muon excess signals from our reconstructed muon data set during the period of
the SPE of 8 November 2000, to see if there also
exist high energy protons of tens of GeV. In view of the features of solar
high energy protons mentioned in Section 1 and 2, and based on our
analysis experience for the GLE of 14 July 2000, we could judge: 1) the GeV high
energy protons, if existed, is more possible coming from a narrow
sky cell; and 2) their onset time to arrive the Earth should be a
little earlier than hundred MeV protons. Observations from  GOES-8
showed the fluxes of protons with energies up to 500 MeV peaked at
about 00:10 UT of 9th November. Thus, the searching for a possible muon excess should mainly focus on
each sky cell and short time period around the peak time of the increase seen by
GOES, starting at 24:00 UT of 8th. That is to see whether there was any time-coincident muon excess with GOES-8 data.

A data set of muons with surface energies over 20 GeV within the
full acceptance of the L3+C detector was used for this analysis.
In order to ensure the real muon events, exact event selections
are necessary for reconstructed muons. In next subsections, we introduce the
event selection criteria and analysis technique which were adopted in previous
study on the GLE of 14 July 2000 \cite{l3sf06,wang09}.

\subsection{Event selection}
As stated above, LEP positron-electron collisor has stopped after the end of October 2000.
But the muon drift chambers of the L3 spectrometer were still running normally. A clean muon data set was obtained.
So in this analysis we did not need to use the cut of variable T$_{0}$ which was applied in previous analysis.
Other previous selection criteria were still
adopted:

1. Only a single muon track is present in the muon chamber;

2. The track is composed of at least 3 segments of hits in
P-chambers (wires parallel to the magnetic field) and by 2
segments of hits in Z-chambers (wires perpendicular to the
magnetic field), ensuring it being a good muon track.

3. The back-tracking of the track from the muon chambers to the
surface is successful in order to ensure good pointing.

\subsection{Time binning and sky mapping }
All selected events were binned in time according to live-time and in space according to
muon¡¯s arrival direction on the ground.

The L3+C data taking system set 0.839 s as a minimal time bin. We also used this minimal time bin as a live-time interval. In our analysis, we combined 100 live-time intervals to form a 83.9 s live-time bin as the
basic time unit.

\begin{figure}
\centering
\includegraphics[angle=0,scale=.50]{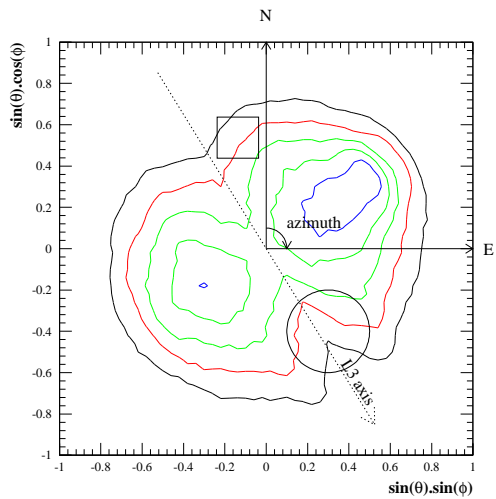}
\caption{The distribution of the arrival
directions of muons observed by the L3+C detector. The contour
lines indicate directions having an equal event rate. The
square represents the sky cell No.48.}\label{cell48}
\end{figure}

In space division the direction cosines\\
\indent \hspace{1cm} $~~l$ = sin$\theta$cos$\phi$\\
\indent \hspace{1cm} $m$ = sin$\theta$sin$\phi$\\
were used as measurables of the muon directions, where $\theta$
and $\phi$ are the zenith and azimuth angles of the muon direction
at the surface. The squared area of the variables $l$ and $m$ was
divided into a $10\times10 (l, m)$ grid. Ignoring those cells with
poor statistics within the detector acceptance, 59 sky cells
containing at least 50 events remained for the investigation. The
contour lines for directions having an equal event rate are shown
in Figure~\ref{cell48} for data of 8 November 2000.

\subsection{Background}
In a day before the flare the Sun was relatively inactive and the GOES satellite showed that the $>$ 4 MeV proton flux was stable \cite{web8}.
So we choose 12 hours data before 21:00 UT of 8th as background measurement.
The same event selection criteria, same time binning and direction
binning were applied to the background analysis.

\subsection{Result}
\begin{figure}
\centering
\includegraphics[angle=0,scale=.50]{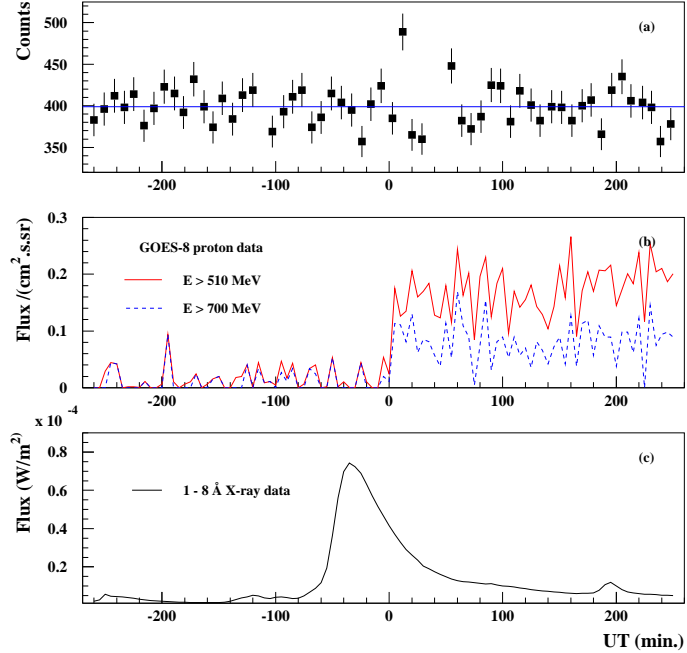}
\caption{a: Number of events versus time in
minutes for 6 hours (around the zero of the horizontal axis, at 24:00 UT of 8th November 2000) in sky cell No. 48.
The live-time bin
width is 8.39 min. The blue solid line shows the mean value of the
background. b: Time profiles of proton fluxes registered by
GOES-8. The red solid line corresponding to proton energies higher
than 510 MeV and the blue dashed line to energies above 700 MeV.
c: Time profile of 1-8 $\AA$ soft X-ray registered by GOES-8.}\label{signal}
\end{figure}

To find possible excesses we compared the data with the background for each sky cell, within the peak time of GOES-8. As a result, an count excess in a bin containing 634
events was found (seeing Figure~\ref{signal}a) in the sky cell No.48 which defined as $0.4375\leq l \leq 0.6375, -0.2375\leq m \leq -0.0375$
(with a solid angle of 0.048 sr). It was within a
8.39 min live-time window (with the real time from 00:07 UT to
00:16 UT of 9th).
This excess was obtained after a first
search for an 83.9 s live-time bin (resulting from the on-line
live-time counting), starting from 24:00 UT and having an
anomalously large number of events followed by other two 83.9 s
live-time bins which also had a higher number of events. The bin
at 00:07 UT that met these requirements was taken as the starting
bin for a possible excess. The following 5 live-time bins were
combined with it to form the 8.39 min live-time window.

\begin{figure}
\centering
\includegraphics[angle=0,scale=.50]{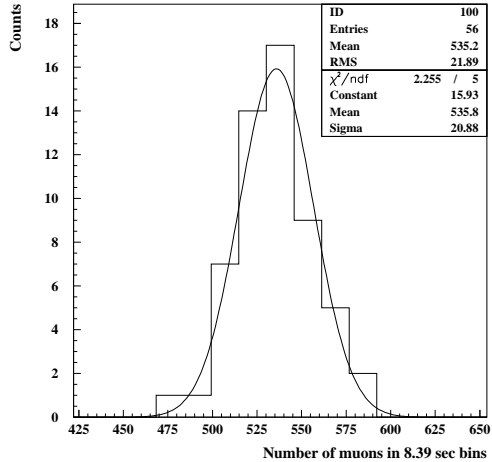}
\caption{The distribution of background events in
the sky cell No.48 obtained from the 12 hours' data before 21:00 UT of 8th November 2000.}\label{background}
\end{figure}

We can see from Figure~\ref{signal}b the excess appeared at a time just coincident with the peak
increase of lower energy solar protons. With 8.39 min
live-time bins we investigated the background of 12 hours before 21:00 UT for the
same sky cell. The background distribution is shown in Figure~\ref{background} and is fitted by a Gaussian.
Using the fitted mean of 535 and the standard deviation equal to
20.9, the excess of 99 events gives rise to a 4.7¦Ò effect.

\begin{figure}
\centering
\includegraphics[angle=0,scale=.50]{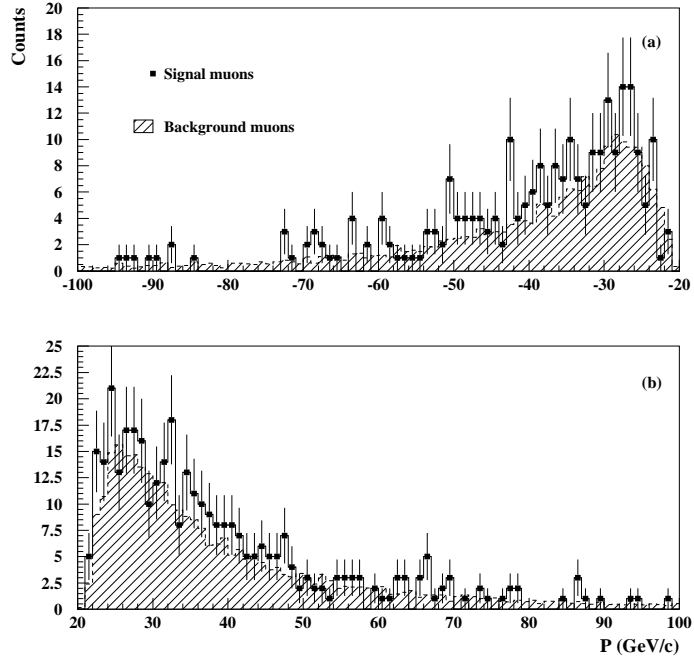}
\caption{Momenta distribution of muons in the cell
No.48 and in the 8.39 min live-time bin at 00:07 UT  of 9th November 2000.}\label{momenta}
\end{figure}

Out of the 634 muons in sky cell No. 48 and in 8.39 min live-time
bin at 00:07 UT of 9th November 2000, there were 373 muons plus and 261 muons minus.
The charge ratio of muons was about 1.43 and the distribution of
their momenta up to 100 GeV/c was shown in Figure~\ref{momenta}.

\begin{figure}
\centering
\includegraphics[angle=0,scale=.50]{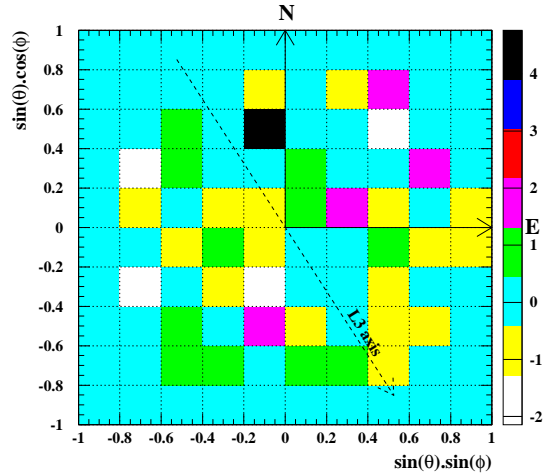}
\caption{The sigma distribution of 59 sky cells in
the 8.39 min live-time bin at 00:07 UT of 9th November 2000.}\label{sigma-map}
\end{figure}

The sigma distribution of 59 sky cells in 8.39 min live-time bin
at 00:07 UT of 9th November 2000 was shown in Figure~\ref{sigma-map}. It is obvious that the most
significant cell appeared in the sky cell No.48.

\section{Discussion}
We have found an excess of 4.7 $\sigma$ in one of 59 sky cells with the selected live-time binning of 8.39 min. The total number of trials is equal to the number of cells timing the number of time window selections. In this searching process the total number of trials was estimated as $59 \times (9+5) = 826$. Here the 9 corresponds to
the number of trials to find the start bin within 00:00 - 00:10 UT
which have large number of events over mean and should be followed
by other two 83.9 s live-time bins which also had a higher number
of events than mean, and the 5 corresponds to the 5 time groups to get
the 8.39 min live-time bin. Based on the total number of trials we can estimate that the probability for such an excess
being due to a background fluctuation is about 0.1\% . Using the
same event selection criteria, same time binning and direction
binning, five days' data (November 5, 6, 7, 10, 11) were
independently analyzed with the `running mean' method. The
result confirmed that no more significant excess than 4.7 $\sigma$
was found except November 8.

Using the air shower simulation code
CORSIKA \cite{Heck1998}, a Monte Carlo simulation
was carried out in order to estimate the primary energies of solar protons which
could be at the origin of the observed excess. The simulation considerations were as following.
Primary protons were
assumed to be incident along the directions that make the produced
muons to appear in the direction of sky cell No. 48. Supposing this
major SPE also having a soft solar proton spectrum like the GLE of
14 July 2000, the index of the primary power law was also set to -6 above 20 GeV.
The simulation result shows that about 90\% of the recorded muons are produced by
primary protons energies ranging from 40 GeV to 200 GeV with a most
probable energy of $\sim$74 GeV. The highest energy of protons is
up to about 1000 GeV. If the primary spectrum is not so steeper,
this value will be higher. Therefore, this observed ¡°excess¡±
could be attributed to solar protons of energies above those a set of NMs being sensitive to.

An upper limit of primary proton flux has also been estimated for this excess.
Sampling a proton flux with a power index -6 and
penetrating into the atmosphere from directions around the sky
cell No.48, muons was produced and traced reaching to the surface by Monte Carlo program.
An area centered around the muon chambers was marked off on the surface. This area should be large
enough to contain the air shower cores, ensuring a very
small fraction loss of muons (less than 1\%). Each muon in this area was traced
through the molasse and the muon chambers and reconstructed using
the same program as for the data. For the background the same
simulation procedure was done except changing a power law index from -6 to -2.7 for a primary cosmic ray
spectrum. Since the primary proton flux is known, we can calculate an upper limit of primary solar protons by comparing the observed data with the simulated data. The flux upper limit of the solar proton
beam entering the upper atmosphere around the direction of the sky
cell No.48 may be

$I(E_p\geq40$GeV)$~ \leq$ ~9.2$\times10^{-3} \rm\;cm^{-2} sec^{-1} sr^{-1}$ (90\% c.l.).

\noindent This value is same order as the one estimated in 2000 July 14
event \cite{l3sf06,wang09}.

\begin{figure}
\centering
\includegraphics[angle=0,scale=.50]{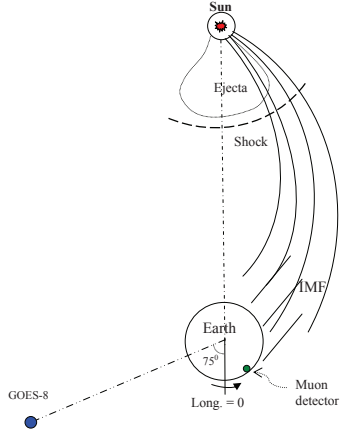}
\caption{Schematic configuration of the relative positions of the
Sun, Earth, satellite GOES-8 and L3+C muon detector at 00:07 UT of 9th November 2000.
The curved dashed lines indicate the ejections and ejection-driven shocks. The Parker spiral field line
connects the Sun to the Earth and its tangent near the Earth along
about 45 degree angle with the Sun-Earth line, which indicates
interplanetary magnetic field (IMF) direction. The prime meridian
at Greenwich (longitude = 0 ) was just at the reverse direction of
the Sun-Earth line.}\label{sky}
\end{figure}

The event of 8 November 2000 is one of the largest SPE except for 16 GLEs
in solar cycle 23. Although not forming a GLE it still possesses following features like most of GLEs: a
related flare locating at more western hemisphere of the Sun, a shock driven by a fast wide CME for particle acceleration,
and a accompanied radio type II/IV burst \cite{wangapp2}.
To help us to understand the particles' transmission and detection, a location map of the Sun, Earth,
satellite GOES-8 and L3+C muon detector just at 00:07 UT of 9th November 2000, ia shown in Figure~\ref{sky}.
In this drawing, the curved dashed thick line represents
shocks driven by large tagged eruptions which expressed as a
dashed thin line. The Parker spiral field line connecting the Sun
to the Earth is drawn and their tangent parallel lines near the
earth form 45$^{0}$ angle with the Sun-Earth line.

It has reported that anti-Sunward detectors may still see high energy solar protons in some extreme events. For example, the earliest arriving particles were detected by stations observing the anti-Sunward hemisphere in the GLE of 28
October 2003 \cite{Bieber2003}. In this event there should exist a great possibility of solar protons being detected by a high directional resolution muon spectrometer, because the in situ detector is not in the position of full back to the Sun, though no NMs had a corresponding.
One reason we speculated may be the following. Since the relativistic solar protons is quite anisotropic, they usually arrive to the Earth in a narrow space direction, that is, forming a solar proton beam. The ground-based NM, as an integrating detector, are usually not sensitive to these particle beams if their flux is not enough large. Even if detected it may be a `hidden registration or poorly-identified' GLE \cite{raa2015}, like a recent SPE on 2014 January 6.

It should be note that the L3 muon spectrometer was still running stably during 2000 November. There was no abnormality in the experimental environment. Furthermore, the background from collision events completely disappeared because LEP positron-electron collisor has stopped running after the end of October 2000. It is because of this background that about 2/3 muon data was cut off in the event of 2000 July 14 \cite{l3sf06,wang09}. Up to now we have not found any other related sources for this muon excess except for time coincident with the solar flare of 8 of November 2000.

\section{Conclusion}
In the solar proton event of 8 November 2000 an excess of 99 muons
with $E_{\mu} \geq$ 20 GeV over a background of 535 was observed in a
particular sky region, lasting from 0:07 to 0:16
UT of 9th. The chance probability for such an excess to be a
background fluctuation is about 0.1\% in this search. It was
time-coincident with the peak increase observed by the satellite
borne detector GOES-8 during the impulsive phase of the solar
flare. If the excess was really induced by solar protons, the
observation indicates that solar protons with energies greater
than 40 GeV were required to produce the excess. If so this may be the second evidence of solar protons above tens of GeV in major SPEs but non-GLEs since 1971.

This study shows that a high directional resolution muon
spectrometer at a shallow depth may detect high energy solar
proton beams which although can not be recorded by NMs. Did it seem to show that a major solar proton non-GLE event
may also produce solar protons with energies above 40 GeV?\\

\noindent {\bf Acknowledgments} We acknowledge the all L3
collaboration members for their unique muon data in this solar
cosmic ray event. We would like to thank Herbert. H. Sauer for
providing GOES data and information about GOES position at the
flare time.

\end{document}